\documentclass[12pt]{article}
\usepackage{graphicx,color}
\usepackage[utf8]{inputenc}
\usepackage{a4,amssymb,amsmath}

\usepackage{newtxtext}

\def\Sy{\mathfrak{S}}

\def\HH{\mathcal{H}}
\def\FF{\mathcal{F}}

\def\ol{\overline}
\def\pa{\partial}

\def\lra{\leftrightarrow}

\def\bea#1{\begin{eqnarray}\label{#1}}
\def\eea{\end{eqnarray}}
\def\ba{\begin{array}}
\def\ea{\end{array}}
\def\bpm{\begin{pmatrix}} \def\epm{\end{pmatrix}}

\def\eref#1{(\ref{#1})}         
\def\sref#1{Sect.~\ref{#1}}

\def\fref#1{Footnote~\ref{#1}}

\def\erw#1{\langle #1\rangle}

\def\wick#1{\,\colon\!\!#1\!\colon}

\numberwithin{equation}{section}
\def\eps{\varepsilon}

\def\sfrac#1#2{\hbox{\large{$\frac{#1}{#2}$}}}

\def\ka{\kappa}

\usepackage{hyperref}
\newcommand\blfootnote[1]{%
  \begingroup
  \renewcommand\thefootnote{}\footnote{#1}%
  \addtocounter{footnote}{-1}%
  \endgroup}

\begin{document}
\title{sQFT: an autonomous explanation of the interactions of quantum particles}
\author{K.-H. Rehren$^1$, L.T. Cardoso$^2$, C. Gass$^3$,  \\ J.M. Gracia-Bond\'ia$^4$, 
  B. Schroer$^5$, J.C. V\'arilly$^6$
\blfootnote{krehren@uni-goettingen.de (corresponding author),
  lucas.cardoso@ufsm.br,  christian.gass@fuw.edu.pl, jose.gracia@ucr.ac.cr,
schroer@zedat.fu-berlin.de, joseph.varilly@ucr.ac.cr}
%
\\[6pt] {\footnotesize $^1$ Institut für Theoretische Physik, 
Georg-August-Universität Göttingen, 37077 Göttingen, Germany.}
\\[-4pt]
{\footnotesize $^2$ Coordenadoria Acadêmica, Universidade Federal de
  Santa Maria,
  Cachoeira do Sul, RS 96503-205, Brazil.} 
\\[-4pt]
{\footnotesize $^3$ Department of Mathematical Methods in
  Physics,
  University of Warsaw, 
  02-093 Warszawa,
Poland.} 
\\[-4pt]
{\footnotesize $^4$ Laboratorio de Física Teórica y Computacional,
Universidad de Costa Rica, San Pedro 11501, Costa Rica.} 
\\[-4pt]
{\footnotesize $^5$ Institut für Theoretische Physik, 
Freie Universität Berlin, 14195 Berlin, Germany.}
\\[-4pt]
{\footnotesize $^6$ Escuela de Matemática, Universidad de Costa Rica, San Pedro 11501, Costa Rica}
}

\maketitle

\begin{abstract}
Successful applications of a conceptually novel setup of Quantum
Field Theory, that accounts for all subtheories of the Standard Model (QED, Electroweak
Interaction and Higgs, Yang-Mills and QCD) and beyond
(Helicity 2), call for a perspective view in a broader conceptual context. The setting is
``autonomous'' in the sense of being intrinsically quantum. Its principles are: Hilbert space, 
Poincar\'e symmetry and causality. Its free quantum fields are
obtained from Wigner's unitary representations of the Poincar\'e
group, with only physical and observable degrees of
freedom. A ``quantization'' of an ``underlying'' classical theory is
not needed.
It allows renormalizable perturbation theory with 
interactions whose detailed structure, and in some cases even the particle content, is predicted by internal consistency. The
results confirm and extend observable predictions for the interactions
of the Standard Model without assuming a ``principle'' of gauge invariance.
\end{abstract}

\hfill {\em Dedicated to Detlev Buchholz on the occasion of his 80th birthday}

\section{The autonomous quantum approach to QFT}
\label{s:aut}

We review ``string-localized quantum field theory (sQFT)'' by
contrasting its conceptual ambitions with standard approaches to perturbative QFT. In later
sections, we shall explain in a largely untechnical way how sQFT proceeds,
and report the results it has produced by now in the context of the
Standard Model of particles (SM).

\paragraph{Conceptual issues of QFT.} Standard textbook approaches to QFT start from classical field
theory plus ``quantization''. At the level of free fields, the strategy
is to use a classical free Lagrangian and ``derive'' canonical commutation
relations that have to be fulfilled in a covariant way. This initial
step is beset with problems: In the case of the Dirac field, the
canonically conjugate momenta are functions of the field itself; and one has
to replace equal-time CCR by CAR ad hoc. In the case of the Maxwell field, one has to
take the unobservable gauge potential as ``canonical'' variables, one has to
change the dynamics by a ``gauge fixing term'' in order to get a
canonically conjugate momentum for all components of the potential, and
one gets CCR that can only be represented on a Fock space with indefinite
inner product (a Krein space), which contains unphysical ``photon states''. We have
learned how to circumvent such trouble by ingenious tricks (Dirac's
theory of constraints, the Gupta-Bleuler method to eliminate
unphysical states with negative or zero norm square, and the
normal-ordering prescription to make the vacuum energy finite). But we
would rather prefer to avoid these first steps altogether.

When it comes to adding an interaction Lagrangian, one usually invokes gauge invariance,
a postulate on unobservable gauge-dependent quantities.
\footnote{
In a famous panel dicussion among P.A.M. Dirac, S. Ferrara,
    H. Kleinert, A. Martin, E.P. Wigner, C.N. Yang, and
    A. Zichichi, recorded in \cite{Zi}, it was consented that gauge theory does
not refer to a ``symmetry'' of Nature. Its main role is the selection of
renormalizable interactions that lie at the basis of the Standard
Model of particles, without constituting a symmetry of the final quantum
theory.} In order to make the previous quantization methods survive the interaction,
in many cases of interest one has to introduce ghost fields and
promote the Gupta-Bleuler condition to BRST invariance. 

Apart from its need of unphysical degrees of freedom in gauge
theories, one may object against ``quantization'' at a more
fundamental level. 
Classical physics is a limiting regime in a quantum world,
  applicable in situations where the uncertainty bound is largely
  exceeded. A prescription to ``read back'' the true theory from its limit precisely in the
  regime where the latter is {\em not} applicable, seems a hazardous
  endeavour. It goes a long way towards explaining the ``need'' for artful but artificial
  devices like the ones just exposed.

The autonomous quantum approach, advocated in this paper, instead starts from quantum principles: Hilbert space, 
unitary Poincar\'e symmetry and Causality. In the words of P. Jordan
in 1929 \cite{Jor29}: it shows how QFT
  can be formulated ``without the crutches of classical correspondences''.
Along the way, it will show that the Wightman axioms (postulating
causality or anti-causality for all Wightman fields, as an extrapolation
from free fields) fall too short for interacting charged fields. sQFT
tells us how interacting charged quantum fields behave instead.

The autonomous approach proceeds as follows. One or several of Wigner's unitary representations of the Poincar\'e
group \cite{Wig} for one's favorite physical particles give rise to a
Fock space, which is a Hilbert space. The
generators of the Poincar\'e symmetry of the one-particle space lift
to the Fock space, and their lifts can be expressed as integrals over 
densities that are already normal-ordered from the
outset. Covariant free fields may be defined with Weinberg's
method
\cite[Chap.~5.]{Wei}, intertwining momentum space Fock space operators with
position space local fields,
and their linear equations of motion follow from the construction. Their
two-point functions and commutators can be computed directly, and causality is automatic for fields associated with particles
of integer spin. (In the massless case, one has to combine helicities
$+h$ and $-h$.)

The spin-statistics theorem is the observation that in the
case of half-integer spin, causality would conflict with the
positivity of two-point functions viewed as inner products of
one-particle state vectors, and the best one can achieve in this case
is anti-locality -- and this forces one to use a fermionic Fock space. 
Equal time (anti-)commutation relations are consequences, and they may not be ``canonical'' at all.
Because anti-locality means maximal violation of causality, fermionic
free fields are not observable fields. Their role is to create fermionic
particle states, and to allow the construction of observable conserved
currents. Their fate when an interaction is turned on has to be
analyzed case by case.

For interactions, one adds an interaction Hamiltonian density, and appeals to
time-dependent perturbation theory with a
spacetime cutoff function $\chi(x)$ to begin with. In the end the
adiabatic limit $\chi(x)\to1$ has to be taken. This limit may be a tough
problem, corresponding to IR problems popping up when one works
without a cutoff function from the outset. The mathematically rigorous
setup for perturbation theory with a cutoff, in which causality is
under control, is known as ``causal perturbation theory'', or
as the ``Epstein-Glaser renormalization scheme (EG)'' \cite{EG}.

Common to both approaches  is the resulting formula for the S-matrix
in the vacuum sector
\bea{STeiL}S= T e^{i\int d^4x\, L_{\rm int}(x)}\eea
(to be properly interpreted), 
where $L_{\rm int}$ is a Wick polynomial in free
fields. The S-matrix is computed perturbatively, by expansion of the exponential. In the
EG approach, one replaces the coupling constant $g$ in $ L_{\rm
  int}(x)$ by a test function $g(x)=g\chi(x)$, 
and regards the resulting ``local S-matrix'' $S[g]$ as an operator-valued
distribution with properties that can be axiomatized (and that may be
established by a perturbative construction).

Worth mentioning is the absence of a ``free Lagrangian
$L_0$'' in this formula
\footnote{Path integral approaches use the
  ``free Lagrangian'' only for the formal definition of a (Euclidean) ``Gaussian
  measure'',
the counterpart of the Fock space whose inner product is the
free two-point function in real time.}, which signals
that its troublesome use for 
the mere purpose of ``quantization'' can (and should) be avoided altogether if
possible. Indeed, the autonomous approach starts from the free fields,
intrinsically defined in terms of the Wigner representations of the
Poincar\'e group, and proceeds with \eref{STeiL}.

In the standard approach, one has to postulate BRST invariance of
the S-matrix, which otherwise cannot be unitary. This is a nontrivial issue, as $\delta_{\rm BRST}L_{\rm
  int}$ is (at lowest order) a total derivative which in general cannot be ignored
because time-ordering does not commute with derivatives, see
\sref{s:sloc}. This failure causes ``obstructions'' against
BRST-invariance of the S-matrix, that have to be taken care of.  
In the form of a ``principle of Perturbative Gauge Invariance (PGI)''
\cite{AS,Sch}, the  BRST method was developed to a mature tool to
predict the structure of interactions, compatible with all sectors of
the Standard Model. It is expanded in G. Scharf's book \cite{Sch}. The
technical (but not the conceptual!)\ aspects of this method are in fact a
precursor of the autonomous approach, which ultimately makes the BRST
method obsolete altogether.

\paragraph{String-localized quantum fields.}
On the Wigner Fock space for helicity 1 (photons), there exist fields that one may
oxymoronically call ``gauge-invariant gauge potentials''. The latter
share formal properties with gauge potentials, but they do not contain 
unphysical modes, see \sref{s:sloc}. 

This miracle is possible because one admits these potentials
to have a weaker localization than
usual: they may be ``localized along a string''. A ``string'' is a
line or a narrow cone extending from $x$ to spacelike infinity, and
``localization'' is understood in the sense of commutation
relations. This feature is the origin of the name ``string-localized
quantum field theory (sQFT)''. String-localized potentials also exist
in the massive case: they have a better UV behaviour than local
massive vector fields, because the weaker localization tames their vacuum fluctuations. This allows power-counting renormalizable
interactions\footnote{We use the term ``renormalizable'' in the sense
  of power-counting throughout. The actual renormalization of sQFT has
not been achieved sofar, but the main result of \cite{G} indicates
that power-counting still is a valid criterion.} that are otherwise
forbidden.

The string-localized potentials are free fields, labelled by some
auxiliary function $c(e)$ characterizing the string, as will be explained in
\sref{s:sloc}. They are used to build  interactions $$L_{\rm int}(c)= g L_1(c) +
\sfrac{g^2}2 L_2(c) + \dots$$ as a Wick polynomial (or possibly a power series) on the physical Hilbert space.  

\paragraph{How sQFT determines interactions.}
(i) $L_{\rm int}(c)$ depends, through the string-localized potentials, 
on the auxiliary function $c(e)$. The point is that the S-matrix must not depend on this
function (``string-independence'').
At lowest order (in the coupling constant $g$), this implies that when $c$ is
varied by a function $\delta c(e)$, then $\delta_cL_1(c)$ must be a total derivative:
\bea{LQ}\delta_c L_1(x,c) = \pa_\mu Q_1^\mu(x,c,\delta c)\eea
by virtue of the equations of motion of the free fields. We call
\eref{LQ} an ``$L$-$Q$ pair''.
The $L$-$Q$ pair condition selects admissible
first-order interactions, to begin with. They are typically cubic in
the free fields. The first noticeable fact is that
all known cubic interactions of the SM are among them, see \sref{s:ex}
for their $L$-$Q$ pairs.

Even more, in all cases of sQFT pertaining to SM interactions, there
is also a point-localized interaction density $K_{\rm int}=gK_1+\frac
{g^2}2  K_2+\dots$ to compare
with. When the Hilbert space where $L_{\rm
  int}(c)$ is defined is suitably embedded into the possibly indefinite state space where
$K_{\rm int}$ is defined, it holds
\bea{LV} L_1(x,c)= K_1(x) + \pa_\mu V^\mu_1(x,c).\eea
We call \eref{LV} an ``$L$-$V$ pair''. It is the point of departure
for an analysis \cite{MRS4,LV} to establish that $L_{\rm int}$ and $K_{\rm
  int}$ produce the same S-matrix, see below.

(ii) Once an initial $L$-$Q$ pair (or $L$-$V$ pair)  is selected, it turns out that the higher orders of the
S-matrix built with $L_1(c)$ will in
general depend on the auxiliary function $c$: there arise ``obstructions against
string-independence'' at higher order of the perturbative expansion.
Their origin, as well as the strategy to cancel
them is quite analogous to the failure and recuperation of BRST
invariance in the PGI approach \cite{Sch}: even if at first order (where there is
no time-ordering) one has
$$\delta_c \int dx \, L_1(x,c) = \int dx\, \pa_\mu Q_1^\mu(x,c,\delta c) =0$$
by virtue of \eref{LQ},
the second-order contribution
\bea{Td}
\delta_c \int dx\, dx' \, T[L_1(x,c) L_1(x',c)] = \int dx\,dx' \, T[\pa_\mu
Q_1^\mu(x,c,\delta c)L_1(x',c) + (x\lra x')]
\eea
will in general not vanish because the derivatives cannot be taken out
of the time-ordered product. It must be cancelled by a second-order
interaction $L_2$, which is only possible if the obstruction has a
suitable form. In that case, it is said to be ``resolved''
by $L_2$, which is called the ``induced'' interaction. Restricting the
analysis of the obstruction to tree level, it follows from Wick's theorem
that if $L_1$ is cubic in the fields, then $L_2$ will be quartic. The second noticeable fact
is that the induced interactions thus determined coincide with the
known quartic interactions of the SM.

(iii) It may happen that obstructions are non-resolvable. A model with
non-resolvable obstructions would have an S-matrix depending on the
auxiliary string variable $c$ without physical meaning, and must be
abandoned as inconsistent with fundamental principles. The third
noticeable fact is that the non-resolvable 
obstructions of several first-order interactions can cancel each other
if all cross-channels are considered. We call this a ``lock-key scenario'': Pieces of the SM
will be inconsistent, while the whole becomes consistent. Examples are
given in \sref{s:ex}.

(iv) The method to compute induced interactions follows a recursive scheme,
whose systematics has been elucidated in \cite{LV}. In principle, it
may continue to arbitrary order. The fourth remarkable fact is that in
all cases investigated so far, the recursion terminates at the
power-counting bound for renormalizability -- without
renormalizability having been imposed.
\footnote{The case of perturbative interactions involving
massless particles of helicity 2 (``gravitons'') is an exception, see \sref{ss:QGC}: the lowest orders of induced
interactions, that have been computed so far, reproduce the
Einstein-Hilbert action which is known to be power-counting
non-renormalizable already at first order, and there is no hint for a termination.}

\paragraph{Interacting fields and observables.}
It is worth remembering that interaction densities $L_{\rm int}$ are just the field-theoretical tool to produce
the corresponding S-matrix of interactions among (asymptotic)
particles, by formula \eref{STeiL}. As long as the S-matrix computed with
string-localized $L_{\rm int}(c)$ is independent of
the string, it appears that one is done. 

One should, however, object that a QFT is more than an S-matrix: it is
defined by its interacting fields whose LSZ limit at $t\to\pm\infty$
is described by that S-matrix. Especially {\em charged} fields are of
interest because they create charged states from the vacuum.
Interacting fields are perturbatively defined by Bogoliubov's formula
\bea{Phiint}\Phi\vert_{L_{\rm int}}(x) = (Te^{i\int L_{\rm int}})^*(T[\Phi(x) e^{i\int
  L_{\rm int}}])
,\eea
to be expanded as a series of retarded integrals over free
fields.

In gauge theory, interacting charged fields are not
BRST invariant and are therefore not defined on the physical Hilbert
space which is defined as the BRST quotient space \eref{HBRST}. In the
BRST approach, one must be content with
an interacting quantum field theory lacking fields of major interest,
that would generate charged states. 

In sQFT,  interacting fields can be constructed by the same formula \eref{Phiint}.
Along with $L_{\rm int}(c)$ being defined on the physical Hilbert
space, so are the interacting fields. The point is, however, that a non-local interaction
density will in general jeopardize locality of the perturbative interacting
fields. Fortunately, there is a powerful
tool in sQFT (the ``magic formula'' \eref{magic} below) that secures
at least string-locality. It proceeds
by relating the interacting fields with interaction $L_{\rm int}(c)$ to
interacting ``dressed'' fields with a {\em local} interaction
$K_{\rm int}$ as in the standard frameworks. The free ``dressed''
fields are much easier to compute than interacting fields, and their
localization can be directly read off. The virtue of the magic formula
is that (because $K_{\rm int}$ is local) the interacting dressed fields have the same localization as
the free dressed fields, and this
determines the localization of the interacting Hilbert space fields
$\Phi\vert_{L_{\rm int}(c)}$ even if $L_{\rm int}(c)$ is string-localized.

Before we shall explain the working of the ``magic formula'', let us pause for a reflection about the meaning of
``localization'' in quantum field theory; and explain how sQFT allows
to assess the localization of interacting fields.

Localization in quantum field theory is an algebraic feature -- not a
geometric one as the term might suggest (and of course not a matter of just
labelling either). An operator is localized
``somewhere'' if it commutes with all observables at spacelike distance
from ``somewhere''. Thus, localization refers to causality of the fields, rather
than locality of the interaction.

A more precise formulation would be that localization is a {\em
  relative} algebraic property because it refers to the localization
of observables. E.g., in perturbation theory, interacting fields are
represented as power series of retarded integrals over free
fields. One may therefore not even expect them to be localized relative to the free fields,
and Borchers \cite{Bo} has shown that they cannot be so if there is nontrivial
interaction.

But they can be localized relative to each other. This property
selects interacting observables. Perturbed fields which fail to be
localized relative to the interacting observables may still be
present in the theory as non-observable fields, and they are welcome
because they create states with non-local
features \cite{BF}. One may call them
``charged fields'' in a generalized sense. Notice that anti-local fields (even free ones)
violate causality in the strongest possible way. This is why only
quadratic expressions like the Dirac {\em current}
$j^\mu=\wick{\ol\psi\gamma^\mu\psi}$ have a chance to qualify as
observables, while the Dirac field $\psi$ creates charged states.

In sQFT, string-localized
{\em free} fields (notably the ``gauge-invariant vector potentials''
mentioned before) arise by integrating a free observable 
along a string emanating from $x$. These appear in $L_{\rm int}(c)$,
and define the perturbation theory. They are of little interest for
the resulting {\em interacting} observables, but they leave their traces on
the interacting charged fields.

\paragraph{The ``magic formula'' settles trouble with non-locality.}
A first reaction on sQFT is often the (in general justified) alarm that a non-local
interaction will spoil locality of the interacting fields. We sketch
here how locality is taken care of when there is an $L$-$V$ pair available.

The  all-orders analysis \cite{LV} of obstructions starting from an initial $L$-$V$ pair
\eref{LV} provides the prerequisites to investigate whether
$K_{\rm int}$ and $L_{\rm int}(c)$ produce the same S-matrix (at
tree level at all orders). This indeed turns out to be the case for all
interactions of the SM. Then, even if $K_{\rm int}$ may be
non-renormalizable, e.g., because it contains couplings of massive vector bosons, the S-matrix $Te^{i\int dx\, K_{\rm
    int}(x)}$ will be unitary and renormalizable -- because $Te^{i\int dx\, L_{\rm
    int}(x,c)}$ is. Conversely, the latter will be string-independent
because the former is.

An offspring of the $L$-$V$ analysis is the ``magic formula''
\bea{magic}
\Phi\big\vert_{L_{\rm int}(c)}(x) = \Phi_{[g]}\big\vert_{K_{\rm int}}(x),
\eea
where the formula to compute the ``dressed fields'' $\Phi_{[g]}$ from
the unperturbed field $\Phi$, using the $L$-$V$ pair, is given in
\cite{LV}. This formula may transfer some of 
the degrees of freedom of the string-localized free fields onto the
dressed fields. The latter may therefore depend on
$c$ and are manifestly relatively local or
string-localized relative to the observable free fields (the latter
can be characterized by the property that $\Phi^{\rm obs}_{[g]}$ is
local -- e.g., if $\Phi^{\rm obs}_{[g]}=\Phi^{\rm obs}$ in the easiest
case). By Borchers'
result \cite{Bo},
this means that $\Phi_{[g]}$ are still not interacting -- but their localization can
be just read off. Then, because $K_{\rm int}$ is a local interaction
density, standard results of causal perturbation theory assure that
$\Phi_{[g]}\vert_{K_{\rm int}}$ have the same {\em relative}
localizations among each other as $\Phi_{[g]}$.
Therefore, $\Phi\vert_{L_{\rm int}}$ is
relatively local resp.\ string-localized to the interacting
observables. This is all one needs to know for the physical
discrimination between ``observable'' or ``charged'' fields. 

Some dressed fields, and consequently also the
corresponding interacting fields are indeed local (and
string-independent), and
qualify as the local observables of the perturbative QFT. Some others
(like the Dirac field of QED) become dynamically string-localized.

This bug is a feature: First, it is unproblematic because charged fields are not
observables. Second, it is even to be expected in theories like QED, because charges can be
measured and have effects at infinite spacelike distance by the Gauss
Law. Notice that the Gauss Law does not violate causality, as the ``creation
of a charge'' is not a physical process. The effects at spacelike
infinity record the behaviour of the charges in the past.

The need to admit string-localized
interacting fields was anticipated by an analysis
(based on the axioms of algebraic quantum field theory referring
exclusively to local observables) \cite{BF} of the causal support where charged
superselection sectors (of massive theories) are distinguish\-able from the vacuum sector. The
result can be interpreted as a statement on putative fields that would
create such sectors from the vacuum, and conforms exactly with the
weaker localization of the interacting fields in sQFT.

\section{String-localized quantum field theory}
\label{s:sloc}

We give here several complementary motivations for string-localized
QFT. The first one starts
from gauge theory and BRST and arises as an attempt to eliminate
all obstructions against BRST invariance, as would appear in the PGI
approach, from the start. Only the next two are truly in the autonomous spirit.
They are enhancements towards applications in the SM of the original
motivation \cite{MSY} which was rooted in Algebraic QFT and
``modular localization''. This background will also be explained briefly below.

For simplicity, and in order to best exhibit the complementarity, we
shall concentrate on the prototypical case of QED, 
with comments on other cases where appropriate.

\paragraph{String-localized quantum fields as the ``reverse side of BRST''.}
If one comes from gauge theory, one would start with a classical
interaction density like
\bea{LQED}L_1(x)=A_\mu(x)j^\mu(x)\eea
for QED.
After covariant quantization, the field $A_\mu(x)$ is
defined on a Fock space that contains states of negative norm square
(``timelike photons'') and states of zero norm (``longitudinal
photons''). One eliminates these unphysical states by the
Gupta-Bleuler method, or equivalently by the more general BRST
method, which is the method of choice in non-abelian gauge theories. The latter 
extends the indefinite Fock space by another indefinite Fock space of
``ghost'' particles. On the extended Fock space $\FF$ there is defined
a nilpotent fermionic BRST operator $Q$ with positive-semidefinite
kernel. Its graded commutator $[Q,\cdot]_\mp$ defines the BRST 
variation $\delta_{\rm BRST}=:s$. Physical states (in QED: photons
with only the two physical helicities, and no ghosts) are those
annihilated by $Q$ modulo states in the range of
$Q$, which are of norm zero. In mathematical language, the physical
Hilbert space  is the quotient space (``cohomology'')
\bea{HBRST}\HH = \mathrm{Ker}(Q)/\mathrm{Ran}(Q).\eea
In order to be defined on this Hilbert space, observables must be
BRST-invariant ($s(X)=0$), including the S-matrix.

One finds that $s(j^\mu)=0$ while
\bea{sA} s(A_\mu) = \pa_\mu u,\eea
where $u$ is a ghost field. Consequently,
\bea{sL}s(L_1) = \pa_\mu u j^\mu = \pa_\mu(u j^\mu)=: \pa_\mu P^\mu\eea
is a total derivative. In particular, $L_1$ itself is not BRST
invariant -- only its classical action integral $\int d^4x\,L_1(x)$ is
BRST invariant. But then the same is not automatically true for the perturbative
S-matrix \eref{STeiL}
because time-ordering does not commute with derivatives. E.g.,
\bea{sTLL}s(T[L_1(x)L_1(x')]) = T[\pa_\mu P_1^\mu(x)L_1(x')]+(x\lra x')
\eea
is in general not a derivative, in which case the integral will not vanish.
This failure is called an ``obstruction'' against BRST invariance. 

Fortunately, QED is particularly good-natured in
that this obstruction actually does not occur, thanks to a Ward identity. (It is particularly
bad-natured by its IR problems, though. The IR aspects of QED will be
addressed separately in \sref{ss:QED}.) But the obstruction does occur in many other
cases of interest, including the
non-abelian case if $L_1$ is only the minimal interaction and the cubic part of the (self-)interaction. The quartic part
of the self-interaction precisely cancels the obstruction. In
\cite{AS,Sch}, the attitude is taken (and proven) that this cancellation mechanism is a
justification for gauge invariant Lagrangians even if one does not assume gauge
invariance from the start.

How would string-localized quantum fields enter the scene?

Because of \eref{sA} and \eref{sL}, one seeks a field $\phi$ whose
BRST variation would be
\bea{sphi} s(\phi) = -u.\eea
Then, one could replace $A_\mu$ by the gauge-equivalent but BRST-invariant field
$A_\mu+\pa_\mu\phi$, and $L_1$ by the BRST-invariant interaction density
$(A_\mu+\pa_\mu\phi)j^\mu $. 

Such a field $\phi$ is usually not ``in the list'' of BRST variations. But
one can easily produce it: the string-localized integral
\bea{IA}\phi(x,e):= \int_x^{\infty}dx'^\mu\, A_\mu(x') = \int_0^{\infty} ds\,
e^\mu A_\mu(x+se)\eea
does the job, with any string direction $e$.

One notices that (by virtue of the homogeneous Maxwell equations)
\bea{IF} A_{\mu}(x,e) := A_\mu(x) + \pa_\mu\phi(x,e) = \int_0^\infty
ds\, F_{\mu\nu}(x+se)e^\nu.\eea
This representation of $A_\mu(x,e)$ makes its BRST-invariance more transparent: it is
just a string integral over the BRST-invariant field
strength. \eref{IF} shows that $A_{\mu}(x,e)$ creates and
annihilates only physical photon states with helicity $\pm1$, both
manifestly (because $F$ does so) and by counting degrees of freedem (namely, \eref{IF} implies $\pa^\mu A_\mu(x,e)=e^\mu A_\mu(x,e)=0$).

Because of the distributional nature of quantum fields,
one must smear the string direction $e$ in \eref{IA}, \eref{IF} with a suitable smooth
function $c(e)$ and a suitable measure $d\sigma(e)$, thus defining $\phi(x,c)$ and
$A_\mu(x,c)$.\footnote{In contrast to the smearing in $x$, we consider
  the smearing
of the string variable as a label of the field $A(\cdot,c)$.} If $\int
d\sigma(e)\,c(e)=1$ (``unit weight''), then
\bea{Ac} A_{\mu}(x,c) =A_\mu(x) +
\pa_\mu \phi(x,c)\eea
is still a potential for $F_{\mu\nu}$:
\bea{dAF}\pa_\mu A_\nu(x,c)-\pa_\nu A_\mu(x,c)=F_{\mu\nu}(x).\eea

One may then work with the BRST-invariant interaction density
\bea{LVQED}L_1(x,c) = A_{\mu}(x,c)j^\mu(x) = A_{\mu}(x)j^\mu(x) + \pa_\mu
\big(\phi(x,c)j^\mu(x)\big)
.\eea
\eref{LVQED} is an $L$-$V$ pair for QED. The S-matrix computed with
$L_1(c)$ will be manifestly
BRST-invariant. In the case of QED, one can establish that the derivative term
does not alter the S-matrix. In more general cases, one may have to add higher-order
interactions, as indicated after \eref{Td}. 

So, it seems that nothing is gained: instead of obstructions against
BRST-invariance, there arise obstructions against string-independence,
that must be resolvable.  But on the contrary: 

Because $L_{\rm int}(c)= gL_1(c)+ \frac{g^2}2L_2(c)+\dots$ is BRST-invariant,
it is defined on the physical Hilbert space \eref{HBRST}, and so will
be the S-matrix and the interacting fields \eref{Phiint}.
One has a (perturbative) QFT on a Hilbert space without unphysical
states from the outset,
including charged fields that can generate charged states from the vacuum, whereas
interacting charged fields in the BRST approach are not defined on the
BRST Hilbert space. See the next paragraph.

\paragraph{The autonomous motivation for massless string-localized quantum fields.}
One can spare the detour through BRST by starting directly from the
integral over $F$ in \eref{IF}
\bea{IcF} A_\mu(x,c):= \int d\sigma(e)\, c(e) \int_0^\infty ds\,
F_{\mu\nu}(x+se) e^\nu
\eea
as an {\em autonomous} definition of the string-localized
potential satisfying \eref{dAF}, where $F$ is defined on the Wigner Fock space.
As in \eref{LVQED}, the interaction is
\bea{Lc}L_1(x,c) = A_{\mu}(x,c)j^\mu(x),\eea
with the distinction that there is no split \eref{Ac} making it an
$L$-$V$ pair, because neither $A$ nor
$\phi$ are defined on the Wigner Fock space. The vital fact is that also without the split, 
$\delta_c(A_\mu(x,c))$ is still the derivative of a quantity defined on the
Wigner Fock space, that we call $w(x,\delta c)$:
\footnote{Because $c$ must have unit weight, a variation must be of
  the form $\delta c(e)= \pa^e_\ka b^\ka (e)$. The field $w(\delta c)$ is then
  computed by integrating $\pa^e_\ka$ by parts onto the $s$-integral
  in \eref{IF} where it produces a derivative $\pa^x_\ka$, and using the homogeneous Maxwell equations. The
  resulting precise form of $w$ is not of interest here. The field
  $Q_1$ contains a factor $w$, and may in more general models than QED also
  involve other string-localized fields.}
\bea{deltaAc}
\delta_cA_\mu(x,c) = \pa_\mu w(x,\delta c).
\eea
Therefore, one has the
$L$-$Q$ pair of QED
\bea{LQQED} \delta_c(L_1(x,c)) =\pa_\mu w(x,\delta c)j^\mu(x) =
\pa_\mu\big(w(x,\delta c)j^\mu(x)\big) =: \pa_\mu Q_1^\mu(x,\delta c)
.\eea
The next steps then follow the familiar scheme: Perturbation theory
with an interaction $L_1(c)$ will in general exhibit an obstruction
against string-independence. In order to resolve the obstruction, one seeks a second-order interaction $L_2$
that makes
\bea{intreso}\delta_c\big( Te^{i\int dx\, (gL_1+\frac{g^2}2 L_2+\dots)}\big) &\!=\!& ig \int dx
\, \delta_c(L_1(x,c)) \\ \notag && \hspace{-15mm}+ \sfrac{(ig)^2}2\Big[\iint dx\,dx'\,\delta_c \big(T[L_1(x,c)L_1(x',c)]\big)
- i \int dx \,\delta_c (L_2(x,c))\Big] + \dots\eea
vanish. The first-order term vanishes thanks to \eref{LQ}. The
condition that the second-order term vanishes at tree level is conveniently
formulated as
\bea{reso}O^{(2)}(x,x')\big\vert_{\rm
  tree}\stackrel!=\delta_c(L_2(x,c))\cdot i\delta(x-x')
+ i\Sy_2 \pa^x_\mu Q_2^\mu(x;x',c,\delta c)
\eea
where $\Sy_2$ is the symmetrization in two arguments, and
\bea{O2}
O^{(2)}(x,x'):=  2\Sy_2\big(T[\pa_\mu Q_1^\mu(x,c,\delta
c)L_1(x',c)]-\pa^x_\mu
T[Q_1^\mu(x,c,\delta c)L_1(x',c)]\big)\eea
is the second-order obstruction that one would get from the first-order interaction alone.
\eref{O2} is analogous (with $Q_1$
in the place of a conserved current) to the violation of a Ward identity. 
By Wick's theorem, its tree-level part can be computed in terms of differences
of time-ordered two-point functions (propagators)
\bea{2po}
O_\mu(\varphi(x),\chi(x')) :=
\erw{T[\pa_\mu\varphi(x)\chi(x')]}-\pa_\mu\erw{T[\varphi(x)\chi(x')]}\eea
of the linear free fields of the model.
The subtractions of explicit derivative terms 
bring the advantage that \eref{2po} and the numerical distributions
in \eref{O2} automatically exhibit delta functions and string-integrated delta
functions.

The existence of $L_2$ and $Q_2$ resolving the obstruction as in \eref{reso}
is a nontrivial feature of
a model. In particular, all string-localized delta functions must be
part of the derivative term $\pa Q_2$. If non-resolvable obstructions
arise, the model must be discarded, unless they can be cancelled (in
the ``lock-key scenario'') by the inclusion of further first-order interactions.
In general, one may have to proceed to higher orders.

The interacting fields $\Phi\vert_{L_{\rm int}(c)}$ are computed with
Bogoliubov's formula \eref{Phiint}. In order to assess their
localization with the help of the ``magic formula'' \eref{magic}, one
embeds the Wigner Fock space
into a suitable indefinite state space
\footnote{\label{fn:embed} In QED, the embedding can be made in
  such a way that the split \eref{Ac} holds on a positive-definite
  subspace with a third degree of freedom besides the two polarizations of the photon.
  Upon interaction, this  extra degree of freedom
is transferred onto the interacting Dirac field -- a feature whose importance has been
stressed in \cite{BCRV}. (The photon continues to have two physical states.)}, where there exists an $L$-$V$
pair with a local interaction $K_{\rm int}$.
 Then, $\Phi\vert_{L_{\rm int}(c)}$ are defined on the Hilbert space, but in general localized in the cone spanned 
by the strings $e$ in the support of $c$. 

In QED, the interacting Maxwell field turns out to be local, while the Dirac
field is localized along the string, see \sref{ss:QED}. This feature resolves the conflict between the
Gauss Law (the total electric charge equals the total electric flux at spacelike
infinity), the Noether property (the total
electric charge operator generates the $U(1)$ symmetry of the charged
field), and locality (the electric flux at infinity should commute with
the Dirac field at $x=0$). There is no such conflict when the last
property is dynamically invalidated. sQFT produces this effect
intrinsically.

\paragraph{The autonomous motivation for massive string-localized quantum fields.}
A second beneficial feature of the formula \eref{IF} arises with
massive vector fields. The Proca field in the Wigner representation
has short-distance scaling dimension 2: in
its two-point function and propagator, $-\eta_{\mu\nu}$ is replaced by
$-(\eta_{\mu\nu}+ m^{-2}p_\mu p_\nu)$. The momentum-dependent term
secures positivity but increases the UV scaling dimension, which is a
measure for the size of vacuum fluctuations. This
jeopardizes the use of the Proca field in perturbation
theory: the minimal coupling to a current has dimension 5 beyond the
renormalizability bound.

One way to deal with this is to just drop the term $m^{-2}p_\mu p_\nu$
which brings down the dimension to $1$. The resulting field is a
local ``massive gauge field'', whose two-point
function violates positivity. One then needs BRST to return to a
Hilbert space. BRST requires gauge invariance, but
the mass violates gauge invariance. So the usual method of choice is to replace
the massive gauge field by a massless one and invoke the
Higgs mechanism to make it ``behave as if it were massive''.

Alternatively (see \cite{Sch}), one may as well introduce a
(local and positive) Stückelberg
field which allows to adapt the BRST method without gauge invariance
and ghosts.
Also in this case, the BRST variation of $L_1$ is a
derivative, causing obstructions of the S-matrix, which can be resolved
only if the vector potential is coupled to a scalar field with the
quartic interaction of the Higgs field.

The autonomous approach, in contrast, secures power-counting renormalizability by
string-localization. The field strength of  the Proca field of dimension
2 also has dimension 2 because the
exterior derivative kills the ``dangerous'' momentum-dependent term in the
two-point function. Then, because the subsequent integration decreases the
dimension, the string-integral \eref{IcF} with the massive
Proca field strength (instead of the Maxwell field strength $F$) is a massive vector
potential of dimension 1 on
the Wigner Fock space of the Proca field. When this field is coupled to a current, the
renormalizability bound is respected. 

For more on this treatment of the massive vector bosons of the electroweak interaction, 
including its ensueing prediction of chirality of the weak
interactions and the shape of the Higgs self-coupling, see \sref{ss:weak}.

\paragraph{Modular localization.} Localization in QFT is an algebraic
feature: observables at spacelike separation must commute with each
other. Usually, one constructs free fields $\varphi(x)$ whose commutator function
is manifestly local. 
Upon perturbation with local scalar interactions, the relative
localization is preserved.

In these approaches, the localization of an operator is encoded in the
test function with which a local field is smeared. In contrast, 
``modular localization'' is a construction of algebras of local
observables that are distinguished by the relevant causal commutativity,
without specific commutation relations when the separation is not
spacelike. It proceeds with Modular Theory.

Modular Theory is an abstract theory about von Neumann algebras $M$ on
a Hilbert space with
cyclic and separating vectors $\Omega\in\HH$. From such a pair $(M,\Omega)$
it allows to extract its ``modular data'': a one-parameter group of automorphisms of $M$ and an anti-linear
conjugation $j$ taking $M$ to its commutant $M'$. These data have
functorial properties which can be exploited for manifold applications
of Modular Theory in QFT (by encoding locality 
in terms of commuting algebras), see \cite{MT}.

``Modular localization''
also starts from the Fock space over unitary Wigner 
representations. On this Hilbert space, one can specify von Neumann algebras $M$ with the 
vacuum as a common cyclic and separating vector, in such a way that their modular data coincide with boost
subgroups of the Poincar\'e group and PCT transformations \cite{BGL}. (For the
latter to be possible, one must combine the massless representations of helicities $+h$ and $-h$.)

Exploiting the correspondence between wedge regions of 
Minkowski spacetime and boost subgroups that preserve these regions,  together with
the functorial properties of modular data, one can {\em consistently define}
algebras of local observables in wedge regions. Namely, the definition complies with the inclusion and
commutation properties that are required for such an interpretation. In other words, the ``localization'' arises from
Poincar\'e symmetry and modular data. Algebras of local observables in smaller regions 
can be defined as intersections of algebras for all  wedge regions that
contain the smaller regions.

Modular localization has not least opened the way to novel non-Lagrangian 
constructions of QFT models with interactions (in two dimensions
sofar), by deformations of von Neumann algebras \cite{Le}.

The method applies as well to the ``infinite spin representations''
of the Wigner classification, for which local free fields do not
exist \cite{Y}. Only in the last step of the  agenda just outlined, the intersections of wedge
algebras defining putative algebras of local
observables in bounded regions, turn out to be trivial. The smallest
intersections of wedge regions that admit nontrivial local observables
are spacelike cones extending to infinity \cite{MSY,LMR}.

This construction suggests that free fields for the infinite spin
representations exist despite \cite{Y}, provided they are allowed to be localized in
cones. Indeed, the authors give concrete examples for such fields in
\cite{MSY}. They then notice that constructions like \eref{IcF} are
possible for every finite spin and helicity, and already observe their
possible usefulness for perturbation theory, as outlined in the previous motivations. 

The present paper intends to show that sQFT lives up to these
expectations.

\paragraph{String-localized quantum field theory in action.}
In \sref{s:ex} we shall look at concrete realizations of sQFT in the Standard
Model of elementary particles. 

A main emphasis will be on the fact that interactions are {\em predicted} by
the need to resolve obstructions, i.e., they are ultimately consequences of
the underlying fundamental principles of quantum field theory:
Positivity, covariance and locality, as explained in
\sref{s:aut}. The strategy is always to find 
first-order (cubic) interactions complying with the condition of
string-independence \eref{LQ}, and study their 
obstructions to determine the higher interactions that
make the S-matrix string-independent by ``resolving the obstructions''  as in \eref{reso}.

Since string-independence is a {\em necessary} condition, it is
legitimate to study the obstructions and their resolution only at tree
level.
Higher-order obstructions at tree level can always be reduced to
expressions of a form generalizing \eref{O2}, which in turn can be computed, with
the help of Wick's theorem, in terms of propagators of the linear free
fields of the model as in \eref{2po}. The systematics at higher orders is fairly simple
when one starts from $L$-$Q$ pairs, and more involved with $L$-$V$ pairs, see
\cite{LV} for details. 

The interactions determined by 
tree-level string-independence are then the starting point for the
UV renormalization of the loop contributions, as in all other
approaches. The power-counting bound being satisfied for $L_{\rm int}(c)$ is
of course instrumental for this step. See \sref{s:outlook}.

The recurrent ``lock-key scenario'' has already been mentioned: The
obstructions of certain first-order interactions cannot be resolved
separately. One has to extend the model by other first-order interactions whose
obstructions cancel each other. E.g., the
minimal interactions in non-abelian models require the cubic
self-interactions of the vector bosons; and non-abelian
self-interactions of massive vector bosons require interactions with a
Higgs field \cite{GRV}. Weak interactions of massive leptons require also Yukawa
couplings to the Higgs field. Notice that these interactions are not
added ``in order to make massless particles massive'', but they are
required by consistency of interactions involving massive particles
from the outset.

In this way, all the interactions otherwise known from gauge theory are
accounted for. The resulting orthodox point of view of sQFT is that one should
abandon ``gauge invariance'' altogether as a physical principle (not
least because it exclusively refers to
unobservable entities). However, we thank D. Buchholz for the comment that also
imaginary numbers are ``not real'', yet they simplify life
enormously. The same seems to be
true for gauge theory methods as compared to sQFT (even if ghost
fields may also be technically awkward).  But the usefulness of gauge
theory should not be mistaken as a ``principle of Nature''.

\section{Examples and achievements so far}
\label{s:ex}

We give a brief overview of the manifold achievements of
sQFT (some of them work in progress). Their relevance for the
Standard Model of elementary particles will become apparent.

We shall present the relevant $L$-$Q$
pairs and some of the resulting obstructions. The treatment of $L$-$V$ pairs 
is technically more intricate, cf.\ \cite{nab}, and will not be
presented here, except for some results concerning dressed fields.

For detailed accounts, we refer to the cited literature.

\subsection{QED}
\label{ss:QED}

The initial $L$-$Q$  and $L$-$V$ pairs of QED were given in
\eref{LQQED}, \eref{LVQED} above. It turns out that 
obstructions of the S-matrix, as illustrated at second order in \eref{O2},
vanish at all orders, and no induced higher-order interaction is
needed.

Among the interacting fields, the Maxwell field tensor and the Dirac
current remain local at all orders. 

On the other hand, the dressed Dirac field is easily computed and turns
out to be string-localized, so that also the interacting Dirac field
is dynamically string-localized. The dressed Dirac field has the
(superficially) simple form
\bea{ddir}\psi_{[g]}(x,c) = \wick{e^{ig\phi(x,c)}}\psi(x),\eea
where $\phi(x,c)$ was defined in \eref{Ac}. The function $c(e)$ must be supported within
the unit sphere of a spacelike plane.
\footnote{The smearing with $c(e)$
is needed because of the distributional nature of quantum fields. It
must be restricted to a spacelike plane in order not to spoil
positivity.} Quantities of the form \eref{ddir} were  previously
considered by pioneers of QED \cite{Jor,Dir,Man} as {\em classical}
expressions with the motive to quantize only gauge invariant quantities.

However, what looks like an innocent classical gauge transformation,
has drastic consequences in quantum theory: the 
exponential of the infrared divergent quantum field $\phi$ turns the
free Wigner Dirac field with a sharp mass into an infrafield which admits no
mass eigenstates: its spectrum is only bounded from below. This
spectral manifestation of the long-distance behaviour of QED was anticipated
in a two-dimensional model \cite{S63}, and later recognized as a
consequence of the Gauss Law of QED in four dimensions \cite{Bu}.

Steinmann \cite{Stein} also used quantum expressions of the form
\eref{ddir} as a starting point to formulate perturbation theory. His idea
is very close to ours, with the distinction that he {\em chose}
\eref{ddir} (not least because it is gauge-invariant), while sQFT {\em
  derives} it.

The quantum field $\phi(c)$ in the exponent \eref{ddir} is IR singular because of the
integration over the massless vector potential involved in \eref{IA}.
But its exponential can be defined non-perturbatively with the help of an IR
regularization \cite{MRS3}; and one finds that the string-smearing function $c(e)$
aquires a physical meaning: it describes the profile of the
asymptotic electric flux density at infinity in the direction $e$,
measured in states created from the vacuum by the field \eref{ddir}.
Loosely speaking, it is the ``shape of the photon cloud'' of an
electron.

This profile is not necessarily uniform in all directions
(because physically it depends on the trajectory of the
electron in the past), and it can be ``engineered'' by choosing the function
$c(e)$ of unit weight. 

States generated by the dressed Dirac field are orthogonal to the
Fock space of the free fields because the coherent state created by
the exponential factor has infinite particle number. In fact, the
cloud function $c$ defines a discrete superselection sector
structure on an uncountably extended Hilbert space such that states
with different $c$ belong to different sectors. More precisely, for
states with several electrons and positrons created by fields with
different $c$, the sum (resp.\  difference) of their cloud functions is superselected.

While this is manifestly true for the free dressed fields, the uncountable
superselection structure survives for the interacting Dirac fields,
but seems to be ``dynamically deformed'' by the interaction $K_{\rm
  int}$ in \eref{magic} \cite{MRS3}.

It was stressed in \cite{BCRV} that it is impossible to satisfy the Gauss Law with
only two photon degrees of freedom, see \fref{fn:embed}. Indeed, unlike the potential $A_\mu(c)$, the field $\phi(c)$ is
not defined on the photon Fock space. By restricting the support of
$c(e)$ to a spacelike plane, its exponential remains positive but contributes one
additional degree of freedom which becomes physical by turning the free Dirac
field into an infrafield \eref{ddir}, and then describes its photon
cloud. The interaction $K_{\rm int}$ then ``hooks'' at the dressing
factor, so that the interacting field complies with the Gauss Law without
``fictitious currents'' (the failure of the free Maxwell equation in
the Feynman gauge by a null field that, however, contributes to the interacting
equations of motion).
The 
field $\phi(c)$ itself is not part of the interacting theory
because of its IR singularity.

The sQFT treatment of the infrared features of QED \cite{MRS3}
encourages a new look at the ``Infrared Triangle'' (soft theorems -- asymptotic
symmetries -- memory effect),
see \cite{Str} for an overview. sQED provides conceptual (in the autonomous sense) elucidations of many of its features.

E.g., sQED allows to compute quantum expectation
values of the Maxwell field in states created by dressed charged
fields, and study their asymptotic behaviour at spacelike ($\mathfrak{i}^0$),
null ($\mathfrak{I}^\pm$), and timelike ($\mathfrak{i}^\pm$)
infinity. 

The behaviour along $\mathfrak{I}^\pm$ can be regarded as a pair of
three-dimensional QFTs with their own intrinsic algebraic structure and
dynamics, arising as asymptotic limits from the bulk, and 
capturing the helicity-one version of the gravitational ``memory effect''. 
The ``matching conditions'' between the past rim of future
null infinity and the future rim of past null infinity can be traced 
by following the continuous interpolation from
$\mathfrak{I}^-$ to $\mathfrak{I}^+$ across spacelike infinity $\mathfrak{i}^0$.

When one views the matching conditions as conservation laws \cite{Str}, the
corresponding infinitely many electric and magnetic conserved
charge operators $Q_\eps$ and $\widetilde Q_\eps$ can be computed, along with the
``large gauge transformations'' that they would generate in the
Feynman gauge. In fact, the latter act also on the additional degree of
freedom in the field $\phi(c)$, emphasized above, 
and consequently on the dressed field and the state it creates from the
vacuum. Thus, while they leave the Maxwell field invariant, they
transform its asymptotic  expectation values. 

The connection with soft photon scattering is manifest in sQED in the
dynamical deformation of the superselection structure \cite{MRS3} caused by the proper
(positivity-preserving) treatment of the IR divergent field $\phi(c)$ in
the exponent in \eref{ddir}.

\subsection{Yang-Mills}
\label{ss:YM}
A cubic $L$-$Q$ pair involving several massless vector potentials
$A^a_\mu(c)$ is necessarily of the form
\bea{YM}
L_1(c) = -\sfrac 12 \sum_{abc} f_{abc}F^{a\mu\nu}
A^{b}_{\mu}(c)A^{c}_{\nu}(c), \qquad
Q_1^\mu(c) = -\sum_{abc}f_{abc}F^{a\mu\nu} w^b(\delta c)A^c_\nu(c).
\eea
A most general ansatz shows that the numerical coefficients $f_{abc}$
must be totally antisymmetric (string-independence at first order) and
must satisfy the Jacobi identity (second order) \cite{GGM}. Thus, they are
necessarily the structure constants of an (unspecified) reductive Lie algebra
$\mathfrak{g}$. (The same result was also found within the PGI approach,
see \cite{AS}.)

If $\mathfrak{g}$ is non-abelian, there is a second-order obstruction
that can be resolved by the induced quartic interaction
$$L_2(c) = -\sfrac{1}2\sum_{abcde} f_{abe}f_{cde} A^a_\mu (c)A^b_\nu(c)
A^{c\mu}(c)A^{d\nu}(c),$$
and higher-order interactions do not
appear. $gL_1(c)+\frac{g^2}2L_2(c)$ is
precisely the string-localized version of the Yang-Mills interaction.

\subsection{QCD}
\label{ss:QCD}
One may add to \eref{YM} an $L$-$Q$ pair of minimal quark-gluon
interactions (copies of \eref{LQQED} with colored currents
$j_a^\mu=\ol\psi \gamma^\mu\tau_a\psi$). The minimal interaction, if taken separately, has a non-resolvable obstruction at second
order:
\bea{nonresQCD}
-\sum_{abc} f_{abc} w^a(x,\delta c)A^b_\mu(x,c)j_c^\mu(x)\cdot\delta(x-x'),\eea
which arises from the violation of the Ward identity for the
non-abelian currents. It is resolved by the same obstruction with the
opposite sign appearing
in the cross contributions when the
self-interaction \eref{YM} is included \cite{nab}. In particular, there is no
induced interaction in the quark sector. This is the first appearance
of the ``lock-key scenario'' mentioned in \sref{s:sloc}. 

The computation of the dressed quark field is much more difficult
than in QED, because of the non-abelianness. We do not have a closed
expression like \eref{ddir}, but the first three orders \cite{nab} indicate 
the onset of a path-ordered exponential
\bea{dqu}\psi_{[g]}(x,c) =\wick{Pe^{ig \phi(x,c)}}\cdot\psi(x).\eea
Recall that $\phi(e)=\sum_aI_e^a(A^a)\tau_a$ is a Lie-algebra-valued line integral over
$A$. The path-ordering orders the Lie-algebra generators, while the
field operators are Wick-ordered. Path-ordering is well defined for a
sharp string ($c(e)$ supported in a single direction) and the
resulting Wilson operator makes \eref{dqu} invariant under {\em
  classical} gauge transformations that are trivial at infinity. For
smooth smearing functions $c(e)$ (which
are needed because of the distributional nature of quantum fields),
``path-ordering'' of the exponential of smeared line integrals is not defined in an obvious way, but \eref{dqu} still
enjoys gauge-invariance (to third order). To give a flavour of the
result \cite{nab}: 
\bea{nabdress}\psi_{[g]}(x,c) =\wick{\exp i \big[g\phi(c) +
\sfrac{g^2}2I_c(i[\phi(c),2A+\pa\phi(c)])  +\sfrac{g^3}6 \phi^{(3)}(c)+
\dots\big](x)}\cdot \psi(x),\quad\eea
where $A= A(c)-\pa\phi(c)$ is the Lie-algebra-valued local potential in Feynman gauge
\footnote{The expression \eref{nabdress} is not manifestly
  positive. But it {\em is} positive, because it was computed with the help of an $L$-$V$ pair on a positive-definite subspace of the
Feynman gauge Fock space, see \fref{fn:embed}.},  $i[\cdot,\cdot]$ is the Lie algebra commutator, and $I_c(Y)$ is
short-hand (appearing already in \eref{IcF}, where we had $\phi(x,c) =
(I_c(A))(x)$) for the string integral operation on a vector field 
$$ (I_c(Y))(x):= \int d\sigma(e)\, c(e)\int_0^\infty ds\, e_\mu
Y^\mu(x+se).$$
The
third-order term in \eref{nabdress} is
$$\phi^{(3)}= 3 I_c(i[I_c(i[\phi,2A+\pa\phi]),A+\pa\phi]) + I_c(i[\phi,i[\phi,3A+\pa\phi]])
-\sfrac32i[\phi,I_c(i[\phi,2A+\pa\phi])].$$
For sharp strings, the nested string integrations of commutators in
the exponent take care of the path-ordering along the string.

As in QED, the string-integrated field $\phi=I_c(A)$ is infrared
divergent, and its appearance in the exponent requires a
non-perturbative definition.  We speculate that analytic confinement
criteria such as \cite{Wil} or \cite{FM} could then be established
already at the level of the dressed quark field \eref{dqu}; but for
this one would need a closed formula at all orders which is presently not available.

\subsection{Electroweak interactions}
\label{ss:weak}
Turning to weak interactions, one may as well generalize the $L$-$Q$
pair \eref{YM} to {\em massive} vector bosons, including self-coupling terms proportional to the masses \cite{GMV,GRV}. One
may again add $L$-$Q$ pairs for minimal couplings to vector or
axial currents $J_a^\mu=c_Vj_a^\mu+c_Aj_a^{5\mu}$. For massive fermions
(because the axial current is not conserved), one also needs couplings
to scalar
and pseudoscalar currents with coefficients determined at second order.

The fermionic sector with the exact field content and masses of the SM
(including the Higgs field) has been analyzed, and its details are
reported in \cite{GMV}. The main emphasis is that the
maximal chirality ($c_A=\pm c_V$) of the weak coupling of the $W$-bosons is a necessary
condition for the resolution of obstructions, which needs both
the  minimal interactions and the self-interactions of the vector bosons.
Chirality is of course well-known as an empirical fact and used as a constitutive
feature of the GSW model, but has no apriori explanation from gauge theory.
Also the empirically known mass-dependence of the Yukawa couplings
follows as a consequence of string-independence at first and second order. 

The bosonic sector is presently under investigation \cite{GRV},
including the self-interaction of massive and massless vector
  bosons and their couplings to the Higgs boson. The
most important feature is that the latter are necessary in
order to resolve non-resolvable obstructions of the
self-interaction. Moreover, the masses $m_W$ and $m_Z$ as the only
empirical input fix, as a consequence of string-independence, the $Z$ and photon
couplings to $W$, and the coupling strengths of $Z$ and $W$
to the Higgs boson. There arise non-resolvable obstructions at third order
in the sectors with one or three Higgs fields. They can
be cancelled if one adds Higgs self-interactions $\ell H^3$ to $L_1$, and
$\ell' H^4$ to $L_2$. The values $\ell$ and $\ell'$ determined by
string-independence are precisely the
same as one also would get from the shifted-symmetric
double-well potential
  $$\sfrac  12 \lambda \big[(\upsilon+H)^2-\upsilon^2\big]^2\quad\hbox{with}\quad g\upsilon=2m_W
  , \quad \lambda =
  m_H^2/4\upsilon^2$$
of the GSW model.
(Details of this latter analysis are almost identical with the abelian model
as reported in \cite{MRS3}.)

Notice that sQFT predicts a Higgs particle along with its
  self-interaction \cite{GRV} -- but
without a ``Higgs mechanism'' because the vector bosons are massive from the start.

Let us point out an interesting pattern of cancellation of
  obstructions, that determines the Lie algebra. It arises most
  cleanly in the
  simpler Higgs-Kibble model, anticipating a similar one in
  the full electroweak theory. The model describes massless fermions
  and vector bosons of equal mass and no photon \cite{nab}.

The $L$-$Q$ pair for the self-coupling is
(with summation over $abc$ understood)
\bea{HKself} \notag
L_{1,{\rm self}}(c) &=& \sfrac 12 f_{abc}\big(F^{a\mu\nu}
A^{b}_{\nu}(c) - m^2 \phi^a(c) 
B^{b\mu}\big)A^{c}_{\mu}(c), \\ \notag Q_{1,\rm self}^\mu(c) &=& f_{abc}\big(F^{a\mu\nu}
A^b_\nu(c) - \sfrac12m^2 \phi^a(c) B^{b\mu}\big)w^c(\delta c)
\eea
where $B_\mu = A_\mu(c)-\pa \phi(c)$ is the local Proca field.

The self-coupling has a non-resolvable obstruction at second order 
of the form
\bea{nonresHKs}
m^2\cdot f_{abe}f_{cde} \cdot w^a(\delta c)A^b_\mu(c)
\phi^c(c)B^{d\mu}.
\eea
This obstruction can be cancelled by another non-resolvable obstruction
arising from the coupling to a scalar field (the Higgs field of mass $m_H$)
with $L$-$Q$ pair (summation over $a$ understood)
\bea{HKhiggs} \notag
L_{1,{\rm Higgs}}(c) &=& \mu\cdot\big(A^a_\mu(c) B^{a\mu}H +A^a_\mu(c) \phi^a(c)\pa^\mu
H  - \sfrac12m_H^2 \phi^a(c)\phi^a(c)H\big), \\ \notag Q_{1,{\rm Higgs}}(c) &=&
\mu\cdot w^a(\delta c)\big(B^{a\mu}H+\phi^a(c)\pa^\mu H\big).
\eea
Its non-resolvable obstruction has the form
\bea{nonresHKh}
-4\mu^2\cdot (\delta_{ab}\delta_{cd} - \delta_{ad}\delta_{cb})\cdot
w^a(\delta c)A^b_\mu(c)
\phi^c(c)B^{d\mu}.
\eea
The cancellation of \eref{nonresHKs} against \eref{nonresHKh} requires
to match the self-coupling coefficients $f_{abc}$ with the Higgs
  coupling $\mu$ via $m^2\cdot f_{abe}f_{cde}=4\mu^2\cdot (\delta_{ab}\delta_{cd} -
\delta_{ad}\delta_{cb})$. This determines the structure constants $f_{abc}=\eps_{abc}$ of $\mathfrak{su}(2)$ and
$\mu=\frac 12m$. (The overall relative factor $\frac12$ as compared to the
abelian model \cite{MRS3} corresponds to the normalization convention of
the generators $\frac12\sigma_a$ of $\mathfrak{su}(2)$.) Other Lie
algebras are presumably possible when nontrivial mass patterns
or several Higgs fields are admitted \cite{GRV}.

\subsection{Helicity 2}
\label{ss:QGC}

In \cite{GGR}, three of us have pursued an analysis of interactions of massless
particles of helicity $\pm 2$ (``gravitons'') with massive or massless matter of spin $0$,
$\frac12$ or $1$. The latter do not need sQFT, but the former do,
because there is no local covariant Hilbert space field to serve as a
``metric deviation'' $h_{\mu\nu}(x)$. There is,
however, a string-localized field $h_{\mu\nu}(x,c)$, built from the
``linearized Riemann tensor field'' on the Fock space over the unitary
Wigner representations of
helicity $\pm2$, by a formula analogous to \eref{IcF} (with two string
integrations).

The analysis of possible cubic $L$-$Q$ pairs results in unique couplings
$\frac12h_{\mu\nu}(c)\Theta^{\mu\nu}_{\rm matter}$ to the conserved
matter stress-energy tensors, 
and in a unique (up to total derivatives) cubic self-coupling. The
matter couplings separately have non-resolvable obstructions, which
are all cancelled in the cross-channels with the self-coupling. This
is another appearance of the lock-key scenario. More remarkably, the
induced interactions (computed only at 
second order) coincide with the classical expansions of 
the Einstein-Hilbert action and of the standard generally covariant
matter couplings, upon substitution of the classical metric deviation
field $h_{\mu\nu}$ by the quantum field $h_{\mu\nu}(c)$.

General covariance was not assumed. It arises
instead as a consequence of the quantum principles underlying sQFT --
a feature that we called ``quantum general covariance''. 

Clearly, the expansion is not power-counting renormalizable. Yet, the
strong constraints underlying the resolution of obstructions may even
raise the hope that the renormalized sQFT of gravity has no free parameters
besides the coupling constant. 

A similar result was obtained in the PGI setting \cite{Dü}: Consistency with
BRST, which in turn is needed to secure positivity, is secured by the expansion of the
Einstein-Hilbert self-interaction in all orders.

\subsection{Beyond the Standard Model?}
\label{ss:BSM}

There are many proposals that ``dark matter'' might consist of
particles of spin or helicity beyond 2.  We
want to discuss what could be possible with sQFT in this respect. 

Recall that by the Weinberg-Witten theorem, particles of helicity $2$ or
higher do not admit local and covariant stress-energy tensor fields on a Hilbert
space. This is often taken as an argument that massless matter of higher helicity 
cannot couple gravitationally, see a discussion in \cite{Porr}. Stress-energy tensors of
massive matter of higher spin exist but have increasing UV scaling dimension,
leading to ``more and more non-renormalizable'' interactions.

On the contrary, there exist string-localized stress-energy
tensors $\Theta^{\mu\nu}_{\rm matter}(c)$ of UV dimension 4 for every
integer spin and helicity \cite{MRS1}. The self-coupling of the
helicity-2 field mentioned in \sref{ss:QGC} is a small modification of 
$\frac12h_{\mu\nu}(c)\Theta^{\mu\nu}_{h=2}(c)$ (which would be part of
an $L$-$Q$ pair only if $\Theta_{h=2}$ were string-independent). This observation raises the hope
that sQFT might also admit $L$-$Q$ pairs for graviton couplings to
massless matter of helicity
beyond 2. Unfortunately, the systematic search turned out no
candidates beyond the known ones with $h=0,1,2$.  The analysis of 
$L$-$Q$ pairs coupling gravitons to particles with half-integer
helicity and to massive higher-spin particles remains to be done.

One may also wonder whether sQFT allows new interactions among known
particles. One such proposal is \cite{Aso}. The authors observed that with an interaction $L'$ involving
the $Z_\mu$ field multiplying two dressing factors $\exp i g \phi(e)$ and $\exp -i
  g \phi(e')$ with $e\neq e'$, one can produce the decay of a $Z$-boson into
two photons, otherwise forbidden by the Landau-Yang theorem. (The
latter uses Lorentz invariance which is manifestly broken by the
choice of the strings.) 

The proposed interaction term $L'$ is not part of an $L$-$Q$ or $L$-$V$ pair, so
the resulting scattering matrix has no reason to be
string-independent. This may not even be intended by the authors;
but what weighs heavier is that as a consequence, there is no ``magic formula'' which
would ensure any kind of localization for quantum fields subjected to an 
interaction $L'$, along the lines discussed in
\sref{s:aut}.\footnote{The authors use an adhoc IR and
UV regularization which in fact violates positivity. Smearing the
strings with $c(e)$ would solve the UV issue. The
positivity-preserving IR regularization used in \cite{MRS2} produces
the superselection structure mentioned in \sref{ss:QED}, and would
actually make the amplitude in \cite{Aso} vanish whenever $c\neq c'$.}

We do not know what is possibly ``allowed'' as
string-localized interactions beyond $L$-$Q$ or $L$-$V$ pairs and higher-order
interactions induced from them, and how much flexibility is gained for
interactions beyond the SM. E.g., $L$-$Q$
pairs that are quartic in the free fields may have slipped our attention. But we strongly feel that one cannot
choose any interaction merely because it seems convenient. Locality and positivity are
precious goods that must be taken care of.

\section{Conclusion and outlook}
\label{s:outlook}

String-localized quantum field theory (sQFT) is a new framework to
address the relation between particles and quantum fields. For free
fields, it is just ``more flexible'' than local QFT. This flexibility
can be used to overcome problems with Hilbert space positivity due to
canonical quantization of massless vector fields, and problems with
renormalizability due to the increased UV scaling dimension of  vacuum
fluctuations of massive vector fields \cite{MRS1}.

sQFT deploys its power when applied to perturbation theory: it strongly
constrains admissible interactions by fundamental
quantum principles (notably Hilbert space). In fact, sQFT can be
understood as a scheme to predict the interactions of the SM in an
``autonomous'' (purely quantum) way.

A prominent lesson that sQFT teaches us is that perturbed quantum
fields, except the observables of the model, must not be expected
to be local (or anti-local). Their dynamical string-localization is a physical
feature (e.g., related to the Gauss Law of QED), and it still allows
``sufficient commutativity'' to conceive a scattering theory in terms
of correlations  at asymptotic times. However, the prevalent methods
of scattering theory have to be properly adjusted.

The dressed Dirac and quark fields \eref{ddir} and \eref{dqu}, that
are (classically) gauge-invariant, are
instances of an issue that calls out for a deeper understanding: given
that sQFT a priori ``knows nothing'' about gauge theory, why do (classically)
gauge-invariant string-localized fields play a distinguished role in sQFT, and why
seem the observables of sQFT to correspond to (classically) gauge-invariant local fields?
Is perhaps ``gauge-invariant and local'' just another characterization
of  ``string-independent''?

An interesting recent argument by Rivat \cite{Riv} proposes
  another reason for ``gauge invariance'': it explains it as a
  consequence of the need to have a Lorentz invariant 
  action $\int d^4x\, L_{\rm int}(x)$. This looks unrelated to sQFT,
  but there is a close connection. Rivat starts from the well-known facts that on the Wigner Fock
space of the photon, a Lorentz covariant vector potential does not exist \cite{Wei};
but one can construct a vector potential $A^{\rm W}_\mu$ which is
Lorentz covariant ``up to an operator-valued gauge transformation'':
\bea{Lor} U(\Lambda) A^{\rm W}_\mu(x) U(\Lambda)^* =\Lambda^\nu{}_\mu \big(A^{\rm W}_\nu(\Lambda x) + \pa_\nu\Gamma(\Lambda,x)\big),
\eea
where $\Gamma$ arises from the ``pseudo-translations'' in the little
group $E(2)$ (the stabilizer subgroup of the Lorentz group of an
arbitrary massless reference momentum vector) in the Wigner
decomposition of $\Lambda$. A transformation law of the same form as
\eref{Lor} holds for
$A_\mu(c)$, where $\Gamma$ arises instead from the Lorentz
transformation of the strings. Consequently, interactions like both $A^{\rm
  W}_\mu j^\mu$ and $A_\mu(c) j^\mu$ change by a total derivative
under Lorentz transformations. 
The main difference between $A^{\rm W}_\mu$ and
$A_\mu(c)$ is the well-controlled localization of the latter.

The focus in this paper was on the relevant analysis to {\em determine}
interactions, which proceeds at
tree-level. The development of an efficient UV renormalization program is the main
open agenda of sQFT. An adaptation of the EG program \cite{EG} to
the string-localized context might be most promising. 

A general strategy can be outlined already. With a fixed smearing
function $c(e)$, string-localized 
two-point functions enjoy the Hadamard property \cite{GPhD}, which allows to
formulate a Wick expansion. Replacing two-point functions by kinematic
propagators, one obtains an unrenormalized time-ordered Wick
expansion, which has to be extended to the singular set where it is
not well-defined. Thanks to the Hadamard property, this set is not
bigger than in the local case \cite{GPhD,G}, and in particular, the usual
power-counting method applies, limiting the scaling degree of the
renormalizations. But there may be more ambiguities in 
the extensions because of the possibility of string-integrated delta
functions. Their classification seems to be the main difficulty, where a
``factorization rule'' for string-localized S-matrized would be
instrumental. Here, ideas from ``string-chopping'' \cite{CMV} are expected
to be helpful. At the same time, the increase of ambiguities will be
(partly) counter-balanced by the condition that string-independence (valid at
tree level) must be preserved.

On the other hand, the equivalence results at tree-level, presented in this
paper and \cite{LV}, 
let us surmise that sQFT will not produce other results than local QFT
with its ``negative probabilities 
plus BRST''. One may therefore do the actual loop calculations in the
usual framework, while the message of sQFT is conceptual: The interactions of
the SM can be understood without gauge fields and ghosts, and are
rather direct consequences of fundamental principles of quantum field theory:
Hilbert space positivity, covariance and locality.

\paragraph{Acknowledgments:} We thank J. Mund and J. Yngvason who
contributed many substantial ideas (some of them unpublished) in
earlier stages of the line of research reported here. We thank
D. Buchholz for interesting and constructive comments. KHR thanks his former MSc and
BSc students,
M. Chantreau, I. Hemprich, T. Joswig, L. Kersten, N. Münch, M. Pabst and F. Tippner, who with
their thesis works have been involved in various explicit 
computations in some of the models of \sref{s:ex} and \cite{nab}. 
CG was funded by the National Science Center of Poland under the grant
UMO-2019/35/B/ST1/01651.

\end{document}